\documentclass[superscriptaddress,groupedaddress,nofootnoteinbib,12pt]{article}  

\usepackage{graphicx}  
\usepackage{dcolumn}   
\usepackage{bm}        
\usepackage{jcappub}

\def\be{\begin{equation}}
\def\ee{\end{equation}}
\def\ba{\begin{eqnarray}}
\def\ea{\end{eqnarray}}
\newcommand{\bem}{\begin{pmatrix}}
\newcommand{\eem}{\end{pmatrix}}
\newcommand{\mk}[1]{\left( #1 \right)}

\newcommand{\f}[2]{\frac{#1}{#2}}

\def\d{\mathrm{d}}
\def\p{{\cal P}}

\def\L*{{\cal L}_*}
\def\L{\mathcal{L}}
\def\({\left(}
\def\){\right)}
\def\ie{{\it i.e. }}

\def\nn{\nonumber}
\def\p{\partial}
\def\mn{_{\mu \nu}}

\def\Ost{Ostrogradsky }
\def\p{\partial}
\def\mupn{^\mu_{\ \nu}}
\def\<{\langle}
\def\>{\rangle}

\def\pa {\partial}

\def\neq {\not\equiv}

\def\cs2{c_{s}^{2}}

\def\p{\partial}

\renewcommand{\thesubsection}{\arabic{section}.\arabic{subsection}}

\begin{document}

\title{Caustics for Spherical Waves}
\author{Claudia de Rham$^{\, a, b}$ and Hayato Motohashi$^{\, c, d}$}
\affiliation{$^a$ Blackett Laboratory, Imperial College London, SW7 2AZ, UK}
\affiliation{$^b$ CERCA, Department of Physics, Case Western Reserve University, 10900 Euclid Ave, Cleveland, OH 44106, USA}
\affiliation{$^c$ Instituto de F\'{i}sica Corpuscular (IFIC), Universidad de Valencia-CSIC, E-46980, Valencia, Spain}
\affiliation{$^d$ Kavli Institute for Cosmological Physics,
The University of Chicago, Chicago, Illinois 60637, USA}
\date{\today}

\abstract{We study the development of caustics in shift--symmetric scalar field theories by focusing on simple waves with an $SO(p)$--symmetry in an arbitrary number of space dimensions.
We show that the pure Galileon, the DBI--Galileon, and the extreme--relativistic Galileon naturally emerge as the unique set of caustic--free theories, highlighting a link between the caustic--free condition for simple $SO(p)$--waves and the existence of either a global Galilean symmetry or a global (extreme--)relativistic Galilean symmetry. }
\maketitle

\section{Introduction and Overview}
\label{sec:Intro}

The notion of causality is anchored at the root of any respectful quantum field theory. In standard Lorentz--invariant theories in flat spacetime, causality if preserved so long as all fields propagate within the same well--defined light--cone. When gravity is included or different notions of metrics come into play these concepts become murkier and much debate still prevails in the literature. \\

In generic theories, the `Characteristics curves' are used to describe the high frequency limit of a signal when the evolution of the background is negligible, see for instance \cite{Felder:2002sv,Babichev:2007dw,Babichev:2016hys,Mukohyama:2016ipl} for relevant discussions in the case of scalar field with non--canonical kinetic terms. For a canonical scalar field in flat spacetime, the characteristic curves represent the field light cone which direction is identical everywhere. In more generic theories, the effective metric of the field may vary from one point to another. In General Relativity it is well--known that the precise opening and tilt of the light--cone may differ from one point to another. In non--canonical theories, the modification in the kinetic term implies that the fluctuations can feel an effective metric which depend on the field background and hence on spacetime. The associated characteristic curves in a non--canonical field theory can therefore differ from straight parallel lines. Even a small shift in these curves may cause them to intersect leading to caustic singularities. See for instance Refs.~\cite{Horie:1999ng,Ehlers:1999bq,Felder:2002sv,Barnaby:2004nk,Harte:2012uw} for more discussions on these effects and on the seminal works of Lax, Jeffrey, Tanuti, Deser, McCarthy and Sario\~glu \cite{Lax:1954gzy,Lax:1957hec,Jeffrey:1964yda,Deser:1998wv} for discussions on caustic--free propagations. \\

In practise, without the existence of a finite UV completion of the theory,  caustic singularities can never be trusted since the second derivatives of the field becomes infinite and the effective field theory leaves its regime of validity. The existence of caustic hence signals the need for new physics to describe the system at that point. Whether and how the new physics involved would resolve the singularity is unknown, but within the effective field theory all one can deduce is that the description breaks down. \\

Nevertheless, even if the caustics could in principle be resolved by appropriate UV completions, their mere existence overcast some shadow on the reliability of some theories. For instance in some models of modified gravity, the creation of caustics has been  shown to be unavoidable for  generic and physically well--motivated initial conditions, with a time scale for the formation which is comparable or smaller than the relevant time scale of system \cite{Contaldi:2008iw,Setare:2010gy}. \\

The formation of caustics was studied in a large variety of modified gravity models and scalar--field theories, including TeVeS (\cite{Bekenstein:2004ne}) in \cite{Contaldi:2008iw}, in Horava--Lifshitz (\cite{Horava:2008ih,Horava:2009uw}) in \cite{Mukohyama:2009tp,Setare:2010gy}, Dirac--Born--Infeld (DBI) in \cite{Felder:2002sv,Goswami:2010rs}.  More recently this effect was investigated for generic scalar field theories with a global shift symmetry, dubbed as  $P(X)$ models, \cite{Babichev:2016hys,Mukohyama:2016ipl} in plane--wave configurations. In these types of theories it was shown that the standard canonical scalar field and DBI are special models for which no caustics form when dealing with simple plane waves. The symmetry of the configurations considered implies that operators with higher derivative without ghost (\ie generalized Galileon operators \cite{Nicolis:2008in,Deffayet:2009mn}) play no role for the evolution and fate of plane waves.\\

When dealing with spherical waves (or as we shall see generic $SO(p)$--waves), the constraints remain exactly identical as that of planar waves for $P(X)$ theories. Beyond $P(X)$ theories, \ie when including generalized Galileon operators,
those operators can play an important role for spherical waves and are constrained by the absence of caustics.  In what follows we show that among all the generalized Galileon operators allowed, only those endowed with a specific additional global symmetry are always caustic--free (as far as spherical waves of arbitrary dimensions are concerned). That global symmetry can be seen to be related to a Poincar\'e invariance in higher dimensions and provides a natural geometric picture from where those theories emerge.  \\

We point out that throughout this analysis we restrict ourselves to a single scalar field theory living on flat space--time.
An important question is whether the flat space--time approximation remains a valid one.  When including interactions with gravity, a single scalar field can be covariantized\footnote{There is some potential freedom in covariantizing the theory without leading to ghost--like instabilities, see Refs.~\cite{Zumalacarregui:2013pma,Gleyzes:2014dya,Motohashi:2014opa,Deffayet:2015qwa,Langlois:2015cwa,Crisostomi:2016tcp,Crisostomi:2016czh,Achour:2016rkg,Motohashi:2016ftl,deRham:2016wji,BenAchour:2016fzp}
for recent discussions as well as Ref.~\cite{Chagoya:2016inc} for a geometrical approach for different covariantizations.} as a generalized covariant Galileon \cite{Deffayet:2009wt} which can be seen to be equivalent to a Horndeski theory \cite{Horndeski:1974wa,Kobayashi:2011nu}. However an alternative perspective is to consider these types of scalar--field models as emerging from the low--energy limit of a infrared modified theory of gravity as proposed in \cite{Nicolis:2008in}. Within that perspective, the `covariantization' of generalized Galileons lie instead with the DGP model \cite{Dvali:2000hr} as shown in \cite{Luty:2003vm}, or within massive gravity \cite{deRham:2010kj} (or its extensions) as shown in \cite{deRham:2010ik}. Within these fully gravitational theories, the additional mode only effectively behaves as a scalar within some low--energy limit, and its behaviour may depart significantly from an isolated scalar at higher energy (as would be the case in the offset of any caustic). This implies that we do expect higher energy effects to kick in on the onset of caustics. The exact effects of gravity is beyond the aim of this work and we restrict ourselves to the case where the fields can be considered to living on flat space--time, for instance by working in an appropriate decoupling limit of the full gravitational theory. The onset of caustics would then signal the breakdown of that decoupling limit.   \\

The rest of the manuscript is organized as follows: In section~\ref{sec:Caustic} we review the characteristic analysis for $P(X)$ theories mainly following the approaches of \cite{Babichev:2016hys,Mukohyama:2016ipl} and extend it to spherical waves of arbitrary dimensions. We then explore the formation of caustics in generalized Galileon theories which affect the propagation of spherical waves in section~\ref{sec:GeneralizedGalileons} and highlight the pure Galileon, the DBI--Galileon, and the extreme--relativistic Galileon, or what we will call the cuscuta--Galileon, which is Galilean generalization of cuscuton~\cite{Afshordi:2006ad,Afshordi:2007yx,Afshordi:2009tt}, as the unique models which are free of caustic formations in arbitrary simple waves configurations.  We further prove in section~\ref{sec:PoincareInvariance} that these theories are the unique ones that satisfy a (non--/extreme--)relativistic Galileon invariance that can be seen as being inherited from a higher dimensional Poincar\'e invariance when considering a probe--brane approximation hence emphasizing the geometrical origin of these special class of models.  After summarizing our results and presenting some outlooks in section~\ref{sec:Outlook}, we provide further details about the simplification of Lagrangians for the DBI--Galileon and the cuscuta--Galileon
obtained from probe--brane approximation, and show that they are included in generalized Galileon in Appendix~\ref{app:probe-brane}.

\section{Caustics and Characteristic Analysis for $P(X)$ theories}
\label{sec:Caustic}

In this section we review the results recently presented in \cite{Babichev:2016hys,Mukohyama:2016ipl} for the formation of caustic in $P(X)$ theories. While those analysis were performed for (simple) plane waves, we show that it is entirely generalizable to any simple wave with $SO(p)$ symmetry where $p\le d$, in a $D=(d+1)$--dimensional spacetime. Consider a $P(X)$ single scalar--field theory in flat spacetime,
\ba
S=\int \d^D x\, P(X)\,,
\ea
with
$X=(\p \phi)^2,$
where all indices are raised and lowered with respect to the flat Minkowski metric.
Obviously, the $P(X)$ theory enjoys the shift symmetry
\be \label{eq:ShiftSym} \phi \to \phi + c , \ee
where the parameter $c$ is constant.
We shall see that imposing caustic--free condition for simple $SO(p)$ waves naturally single out theories that enjoy additional global symmetry.

\subsection{$SO(p)$--waves}
\label{sec:SOp}

We now consider waves with an $SO(p)$ symmetry, so that the field only depends on the coordinates $t$ and on the distance $r=(x_1^2+\cdots+x_p^2)^{1/2}$ in the $p$-dimensional subspace. The case of planar symmetry considered in \cite{Babichev:2016hys,Mukohyama:2016ipl} therefore corresponds to the special case where $p=1$. Assuming the $SO(p)$ symmetry, we then have
$X=-\dot \phi^2+\phi'{}^2,$
where dots represent derivatives with respect to $t$ and primes represent derivatives with respect to the variable $r$.\\

The equations of motion for $\phi$ are hence simply given by
\ba
\label{eq:EOMPX}
Z^{\mu\nu}\p_\mu \p_\nu \phi+\frac{p-1}{r} \phi' P'(X)=0\,,
\ea
where $Z^{\mu\nu}$ is conformally related to the effective metric and independent on $p$,
\ba
\label{eq:ZuvPX}
Z^{\mu\nu} &=& P'(X)\eta^{\mu\nu}+2P''(X) \pa^\mu\phi\pa^\nu\phi \notag\\
&=&
\bem
-P'+2P''\dot\phi^2 & -2P'' \dot\phi \phi' \\
-2P'' \dot\phi \phi' & P'+2P''\phi'{} ^2
\eem \,.
\ea
The two eigenvalues of $Z\mupn$ are $P'(X)$ and $2XP''(X)+P'(X)$ and so the speed of sound is given by
\ba
\label{eq:csPX}
c_s^2=\frac{P'(X)}{2XP''(X)+P'(X)}\,.
\ea

It is clear from the equations of motion \eqref{eq:EOMPX} that the only difference between a plane--wave ($p=1$) and other $SO(p)$--wave is in the contribution from the second term in \eqref{eq:EOMPX} which only involves first order derivatives on the field. As a result the parameter $p$ cannot affect the causal structure of the field and neither can it change when caustic can be generated. It therefore directly follows that all the results derived in \cite{Babichev:2016hys,Mukohyama:2016ipl} in the case of $p=1$ are directly applicable for arbitrary $p$. However one significant difference between planar waves and waves involving more spatial directions is that the Galileon interactions are only relevant for $p>1$ and by looking at waves beyond the planar symmetry one can also isolate special types of Galileons interactions that are caustic--free for these types of waves. Before including the Galileon interactions, it is useful to establish how the argument for the generation of caustic applies in the case of $P(X)$ theories and to establish the connection with a characteristic analysis.

\subsection{Characteristic analysis}
\label{sec:charac}

The characteristic analysis focuses on the highest derivative terms, which are therefore the  second derivatives in the equation of motion \eqref{eq:EOMPX}. As a result, the parameter $p$ is therefore irrelevant in this analysis. First we notice that at high energy, the stability condition imposes $\det Z <0$, we therefore restrict ourselves to this case in what follows.\\

Then focusing on the high energy terms, we can use the substitution $\p_\mu \to k_\mu$  and at high energy the vector $k_\mu$ is parallel to:
\be \label{psols}
k_\mu^{(\pm)}=A(t,r)
\left\{\begin{array}{ccl}
\mk{ \xi_\pm  ,-1}, \quad & \qquad{\rm if}  & \quad Z^{00}\neq 0, \\
\mk{-1,\f{Z^{00}}{Z^{11}}\xi_\pm},  & \qquad{\rm if}& \quad Z^{11}\neq 0, \\
\mk{ -1, \f{2Z^{01}}{Z^{11}} } ,~(-1,0),  & \qquad{\rm if}& \quad Z^{00} = 0, \\
(0,-1),~ \mk{ -1, \f{Z^{00}}{2Z^{01}} },   & \qquad{\rm if}&  \quad Z^{11} = 0\,,
\end{array}\right.
\ee
where $A(t,r)$ is an overall factor, and we have used the notation,
\ba
\label{xipm}
\xi_\pm=\f{Z^{01}\pm \sqrt{-\det Z} }{Z^{00}}= \f{v\pm c_s}{1\pm v c_s}\,,
\ea
with $v=-\frac{k_r}{k_t}$ and with the speed of sound $c_s$ given in Eq.~\eqref{eq:csPX}.
The overall factor $A(t,r)$ is determined by the integrability condition $\p_t k_r=\p_r k_t$, with which one guarantees the existence of an integral function $\sigma_{\rm}(t,r)$ satisfying
\be
\pa_\mu \sigma_\pm = k_\mu^{(\pm)}\,.
\ee

The constant-$\sigma_\pm$ surface defines the characteristic surface (or curve in $1+1$ dimension), whose tangent vector given by $Z^{\mu\nu}k_\nu^{(\pm)}=\pm\sqrt{-\det Z}(1,\xi_\pm)$ is the direction of propagation.
Since the light cone is defined by $Z^{\mu\nu}k_\nu^{(\pm)}$, it is sufficient to analyse the vectors $\(1, \xi_\pm\)$ to determine the causal structure of the theory and in particular establish the existence of caustics.\\

In terms of $\xi_\pm$, the equation of motion \eqref{eq:EOMPX} for the scalar field $\phi$ can be written as
\ba
\p_{\sigma_\pm}k_t+\xi_\mp \p_{\sigma_\pm}k_r=0\,,
\ea
or
\ba
\(\frac{\d k_t}{\d k_r}\)_\pm=-\xi_\mp\,,
\ea
and one can use the expression \eqref{xipm} for $\xi_\pm$ in terms of the speed of sound to determine the speed of sound for a given $\xi$,
\ba
\label{cs xi}
c_s^2=\(\frac{v-\xi}{1-v \xi}\)^2\,.
\ea

Following the analysis of Ref.~\cite{Mukohyama:2016ipl}, one can consider $Z^{00}\neq 0$ (while still focusing on $\det Z<0$ for stability reasons as mentioned previously). The case where $Z^{00}=0$ can be considered separately but does not lead to any different result as explained in \cite{Mukohyama:2016ipl}. Then focusing on the first solution in \eqref{psols}, one can take $k_\mu^{(\pm)}\sim \mk{ \xi_\pm  ,-1}$. A vector orthogonal to $k_\mu^{(\pm)}$ is given by $(1,\xi_\pm)$, which corresponds to $\pa_{\sigma_\pm} \propto \pa_t + \xi_\pm \pa_r$ in~\cite{Mukohyama:2016ipl}, and precisely coincides with the direction of the vector $Z^{\mu\nu}k_\nu^{(\pm)}$.  The surfaces tangent to  $\pa_{\sigma_\pm}$ then define the $C_\pm$-characteristics.

\subsection{Caustics}

Generically, caustics arise when two (or a set of) $C_+$-characteristics or $C_-$-characteristics are able to intersect (see Refs.~\cite{Felder:2002sv,Babichev:2007dw,Babichev:2016hys,Mukohyama:2016ipl,Horie:1999ng,Ehlers:1999bq,Barnaby:2004nk,Harte:2012uw,Lax:1954gzy,Lax:1957hec,Jeffrey:1964yda,Deser:1998wv}). At those points where characteristics converge, the first derivative of the field is not well--defined and its second derivative diverges, signaling the breakdown of the effective field theory. In an EFT approach, caustics should not be understood as real singularities since they `only' signal the need for new physics to provide the correct description of the system at those points. However one  could still raise the question of whether there exists systems for which the absence of caustics is automatically guaranteed (at least for simple waves). For arbitrary waves, or arbitrary spacetime backgrounds, the answer is likely to be uniquely restricted to canonical fields \cite{Babichev:2016hys}, however restricting ourselves to $SO(p)$ waves, and particularly simple waves, we see that DBI takes on a special place.  \\

Indeed, asking for the absence of caustics to be guaranteed for simple $SO(p)$ waves is equivalent to asking for the characteristics to be parallel in the $(k_t, k_r)$ plane \cite{Mukohyama:2016ipl}, \ie requiring the existence of constant coefficients $a$ and $b$ so that
\ba
k_t=a\, k_r + b\,.
\ea
As proven in Ref.~\cite{Mukohyama:2016ipl}, this linearity condition is actually equivalent to the exceptional condition pointed out in \cite{Deser:1998wv}.\\

We can therefore infer that
\ba
X=-k_t^2+k_r^2=(1-a^2)k_r^2-2 a b k_r-b^2\,.
\ea
From the relations $v=-k_r/k_t$ and $\xi_\mp = - \d k_t / \d k_r = -a$ we can use the expression \eqref{cs xi} to deduce that the absence of caustics implies that the speed of sound should be linear in $X$:
\ba
\label{eq:csab}
c_s^2 (X) = 1+\alpha X\,,
\ea
with
\be \label{eq:defal} \alpha \equiv \f{g}{f} \equiv \f{1-a^2}{b^2}. \ee
Using rescaling the field, without loss of generality we consider the following cases: $(f,g)=(\pm1,0),(1,\pm 1),(0,\pm 1)$, namely, $\alpha=0,\pm 1,\pm \infty$. \\

We can now compare this expression for the speed of sound with the one given in \eqref{eq:csPX} for a $P(X)$ model and infer the following differential equation for $P(X)$ so as to avoid caustics,
\ba
\label{eq:diffP}
\frac{\d}{\d X}\log P'(X)=\frac{1}{2X}\(\frac{1}{c_s^2(X)}-1\)=-\frac{\alpha}{2}\(1+\alpha X\)^{-1}\,.
\ea
This analysis was derived in \cite{Mukohyama:2016ipl} and we simply reproduce it here to simplify the discussion when including Galileons. Notice however that once again, this result is not only valid for simple plane waves, but also for simple spherical waves, cylindrical waves or any simple wave with a $SO(p)$--symmetry. In other words, simple spherical and cylindrical waves do not impose any additional conditions compared to plane waves as far as caustics are concerned in $P(X)$ models. \\

Demanding that for any $b$ there exists an $a$ such that the differential equation \eqref{eq:diffP} be identically satisfied for any $X$  (or vise--versa for $a$ and $b$),
we infer that there are only three possible relevant systems, namely, the canonical scalar field for $\alpha =0$ (or $(f,g)=(1,0)$), DBI for finite values of $\alpha$, and the cuscuton for infinite $\alpha$.  Indeed we obtain
\ba
\label{eq:P(X) no caustic}
P(X)=\left\{\begin{array}{ccl}
X & & {\rm for } \  \alpha =0, \  ({\rm or }\ (f,g)=(\pm 1,0) )\\[10pt]
\sqrt{ 1 \pm X }& & {\rm for } \  \alpha = \pm 1, \ ({\rm or }\ (f,g)=(1,\pm 1) ) ,\\[10pt]
\sqrt{\pm X}& & {\rm for } \  \alpha \to \pm \infty, \ ({\rm or }\ (f,g)=(0,\pm 1)) .
\end{array}\right.
\ea
The pure Galileon $P(X)=X$ and DBI model $P(X)=\sqrt{1 \pm X}$  appear as the unique theory among generic $P(X)$ models \cite{Mukohyama:2016ipl} from the caustic formation point of view.  In addition, when taking the extreme limit of DBI ($\alpha \to \infty$) we recover the cuscuton model $P(X)=\sqrt{\pm X}$ introduced in~\cite{Afshordi:2006ad,Afshordi:2007yx,Afshordi:2009tt}, whose speed of sound is infinite and kinetic term reduces to a total derivative for homogeneous configuration in flat spacetime. In the case of the cuscuton, the theory makes sense for either a timelike or a spacelike field. \\

\subsection{Additional global symmetries}
\label{sec:addsym}

Interestingly, the three caustic--free models \eqref{eq:P(X) no caustic} enjoy the additional global symmetry apart from shift symmetry.
First, DBI is of course also well--known to be a privileged class within generic $P(X)$--models for other reasons. Indeed in addition to the shift symmetry it enjoys an addition global symmetry. The DBI action is invariant under the following transformation $\phi \to \phi+\delta \phi$, with
\ba
\label{eq:DBIsym}
\delta \phi= c+ v_\mu x^\mu + \phi v^\mu \p_\mu \phi\,,
\ea
where the parameters $c$ and $v_\mu$ are constant.
Indeed, under this transformation the DBI Lagrangian transforms as a total derivative,
\ba
\delta \(\sqrt{1+X}\) \sim \frac{\delta X }{\sqrt{1+X}} \sim  \frac{v^\mu \p_\mu \phi  (1+X) + \phi v^\mu \p_\mu \p_\nu \phi \p^\nu \phi }{\sqrt{1+X}}\sim v^\mu \p_\mu \(\phi \sqrt{1+X}\)\nn\,,
\ea
and the DBI action is thus invariant.
DBI can be seen to arise as the motion of a probe--brane in a higher dimensional spacetime. Within that picture, the previous symmetry \eqref{eq:DBIsym} is simply reminiscent of invariance under the higher--dimensional rotations and boosts.\\

On the other hand, the pure Galileon $P(X)=X$ and the cuscuton $P(X)=\sqrt{X}$ enjoy slightly different form of the symmetry, which are some limit of \eqref{eq:DBIsym} from higher--dimensional point of view.
Indeed, it is well--known that the pure Galileon $P(X)=X$ enjoys the global symmetry
\ba
\label{eq:GalileanSym}
\delta \phi= c+ v_\mu x^\mu ,
\ea
which can be viewed as non--relativistic limit of the transformation \eqref{eq:DBIsym} in higher--dimensional description.
Similarly, one can check that the cuscuton model is invariant under
\ba
\label{eq:exGalsym}
\delta \phi= c+ \phi v^\mu \p_\mu \phi\,,
\ea
which amounts to extreme--relativistic limit of the transformation \eqref{eq:DBIsym}.  Therefore, all three caustic--free classes \eqref{eq:P(X) no caustic} enjoy the additional global symmetry, which is inherited from a higher--dimensional description. \\

While the exact link is not entirely flushed out, it is very likely that the higher--dimensional origin of these symmetries is the reasons why the models are protected against caustics, at least as far as simple $SO(p)$--waves are concerned. With this in mind, it is therefore likely that other theories which also enjoy a natural higher--dimensional geometrical probe--brane origin could be protected against the same type of caustics. In what follows we will therefore study the emergence of caustics in generalized Galileon theories (as derived in \cite{Deffayet:2009mn}), and establish that the pure Galileon, the DBI--Galileon, and a Galilean generalization of the cuscuton dubbed the cuscuta--Galileon, which also come from a similar higher--dimensional description, satisfy the same properties as models \eqref{eq:P(X) no caustic}, at least as far as caustics for simple $SO(p)$--waves are concerned. \\

Before moving to generalized Galileons, it is worth mentioning that while DBI appears to be special with respect to caustic formation, when it comes to quantum corrections, the global symmetry \eqref{eq:DBIsym} is not sufficient to fully protect the theory (unless the quantization prescription also makes use of the symmetry). See \cite{deRham:2014wfa} for more details. \\

We also emphasize that given DBI has a speed of sound $c_s^2=1+X$, we can automatically infer that there are always configurations where the model is superluminal. Indeed while the speed is subluminal for a timelike field ($X<0$), the speed is always superluminal for a spacelike field ($X>0$). Had we chosen a negative $\alpha$, we would have observed the opposite.
On the other hand, since $\alpha \to \pm \infty$, so long as $X\ne 0$, the cuscuton has an infinite speed of sound, or is incompressible, which leads to an instantaneous propagation.
Yet, the superluminalities and instantaneous propagation do not necessarily automatically imply the existence of closed timelike curve or acausalities (see \cite{Milonni,Brillouin,Visser:1992py,Bruneton:2006gf,Bruneton:2007si,Babichev:2007dw,Hollowood:2007kt,Shore:2007um,Burrage:2011cr,deRham:2014zqa,Motloch:2015gta,Motloch:2016msa,Afshordi:2006ad,Afshordi:2007yx} for more discussions).

\section{Caustics in Generalized Galileons}
\label{sec:GeneralizedGalileons}
\subsection{Generalized Galileons}

We now extend the analysis to include the generalized Galileon operators \cite{Nicolis:2008in,Deffayet:2009mn} given in $D=d+1$--spacetime dimensions by
\ba
\label{eq:Gal}
\label{eq:Lnapp}
\L_{n}=G_n(X)\,  \delta^{\mu_1 \cdots \mu_{n-2}}_{\nu_1 \cdots \nu_{n-2}} \, \Phi^{\nu_1}_{\mu_1}\cdots \Phi^{\nu_{n-2}}_{\mu_{n-2}}=G_n(X)\,  \mathcal{G}_{n-2}[\Phi]\,,
\ea
with $n=2,\cdots,D+1$, $\Phi\mn=\p_\mu\p_\nu \phi$ and
where we use the notation
\ba
\delta^{\mu_1 \cdots \mu_{k}}_{\nu_1 \cdots \nu_{k}} &=&\mathcal{E}^{\mu_1 \cdots \mu_D}\mathcal{E}_{\nu_1 \cdots \nu_D}  \, \delta^{\nu_{k+1}}_{\mu_{k+1}}\cdots \delta^{\nu_{D}}_{\mu_{D}}  \\[5pt]
\mathcal{G}_k[\Phi]&=& \delta^{\mu_1 \cdots \mu_{k}}_{\nu_1 \cdots \nu_k} \, \Phi^{\nu_1}_{\mu_1}\cdots \Phi^{\nu_{k}}_{\mu_{k}} \,,
\ea
for $0\le k \le d$ and
where $\mathcal{E}_{\mu_1\cdots \mu_D}$ is the Levi-Civita tensor.\\

It will also be convenient to introduce the functional tensor $X^{\mu\nu}_n$ defined as \cite{deRham:2010ik,deRham:2010kj,deRham:2014zqa}
\ba
\label{eq:tensorX}
X^{\mu\nu}_{n}[\Phi]=\frac{1}{n+1}\frac{\delta }{\delta \Phi\mn}\mathcal{G}_{n+1}[\Phi]
= \eta^{\nu\beta}\delta^{\, \mu\, \alpha_1\cdots \alpha_n}_{\, \beta\, \beta_1\cdots \beta_n}\Phi^{\beta_1}_{\alpha_1}\cdots \Phi^{\beta_n}_{\alpha_n}
\,,
\ea
for $0\le n \le d$, so that
\ba
\mathcal{G}_n[\Phi] = \Phi_{\mu\nu}X^{\mu\nu}_{n-1}[\Phi]\,.
\ea
In particular we have $X^{\mu\nu}_{-1}[\Phi]=0$, $X^{\mu\nu}_0[\Phi]= \eta^{\mu\nu}$, $X^{\mu\nu}_1[\Phi]=\Box \phi \eta^{\mu\nu}-\Phi^{\mu\nu}$ and the symmetric  tensor  $X^{\mu\nu}_n$ satisfies the following useful properties (see Refs.~\cite{deRham:2010kj,deRham:2014zqa})
\ba
\p_\mu X^{\mu\nu}_n&=&0, \quad \forall \ n\ge0\\
X^{\mu\nu}_n[\Phi]&=&-n\, \Phi^{\mu}_{\alpha}X^{\alpha \nu}_{n-1}+\Phi_{\alpha\beta}X^{\alpha\beta}_{n-1}\eta^{\mu\nu},\quad \forall \ 1\le n\le D-1\,. \label{eq:recX}
\ea\vspace{5pt}

To maintain the shift symmetry,  we restrict ourselves to functions $G_n$ that only depend on $X$ and not on the field $\phi$ itself.
For concreteness, we note that $\L_2$ is nothing other than a standard $P(X)$ model, $\L_3$ corresponds to the generalized cubic Galileon, etc.,
\ba
\L_2&=&G_2(X) \,,\\
\L_3&=&G_3(X)\Box \phi\,,\\
\L_4&=&G_4(X) [(\Box \phi)^2 - (\pa_\mu\pa_\nu\phi)^2] \,,\\
\L_5&=&G_5(X) [(\Box \phi)^3 - 3\Box\phi(\pa_\mu\pa_\nu\phi)^2 + 2 (\pa_\mu\pa_\nu\phi)^3] \,,
\ea
up to an irrelevant overall dimensionless constant. For arbitrary functions $G_n(X)$ the theories only enjoy a shift symmetry $\phi\to \phi+c$,
which is broken if an explicit $\phi$ dependence is introduced in $G_n(\phi,X)$.\\

We recover the standard Galileon interactions when  the functions $G_n(X)$ take the particular form $G_n(X)=X$, and these are the unique set of interactions (without an \Ost ghost) which enjoy an additional non--relativistic Galilean global symmetry \eqref{eq:GalileanSym} \cite{Nicolis:2008in}.
If $G_n(X)$ takes on another very particular form, we shall see that the theory enjoys instead the relativistic Galilean symmetry \eqref{eq:DBIsym} or the extreme--relativistic Galilean symmetry \eqref{eq:exGalsym}.

\subsection{Equations of Motion for $SO(p)$--wave}

We now consider a configuration with an $SO(p)$--symmetry as in section~\ref{sec:SOp}, where the field solely depends on the time $t$ and on the $p$--dimensional distance \mbox{$r=(x_1^2+\cdots x_p^2)^{1/2}$}, where $p$ is an arbitrary integer with $0\le p \le d$. Under those configurations, the generalized Galileon interactions for $1<n<p+3$ then take the form
\ba
\label{eq:LnSOp}
\L_{n}= G_n(X)\Bigg[&& \hspace{-12pt}\frac{(p-1)!}{(p+1-n)!}\(\frac{\phi'}{r}\)^{n-2}
+\frac{(n-2)(p-1)!}{(p+2-n)!}\(\frac{\phi'}{r}\)^{n-3}\(\phi''-\ddot \phi\)\nn \\
&+&\frac{(n-2)(n-3)(p-1)!}{(p+3-n)!}\(\frac{\phi'}{r}\)^{n-4}\(\dot \phi'{}^2-\ddot \phi \phi''\)\Bigg]\,.
\ea
We directly see as expected from the symmetry that the Lagrangian identically vanish for $SO(p)$ waves with $p<n-3$, and is a total derivative for $p=n-3$. This naturally explains why no Lagrangian with $n\ge 3$ is relevant to the study of plane waves with $p=1$, but start becoming relevant for more general $SO(p)$ waves.\\

The equations of motion for the generalized Galileon can then be written in the form $\sum_n \mathcal{E}_{n}=0$, with
\ba
\label{eq:eomGal}
\mathcal{E}_{n}&=&
\frac{(p-1)!}{(p-n)!}\(\frac{\phi'}{r}\)^{n-1}G_{n, X}+
\frac{(p-1)!}{(p+1-n)!}\(\frac{\phi'}{r}\)^{n-2}Z_n^{\mu\nu}\p_\mu \p_\nu \phi\\
&+&\frac{(n-2)(p-1)!}{(p+2-n)!} \left[(n-1)G_{n,X}+2 G_{n,XX}X\right]\(\frac{\phi'}{r}\)^{n-3}\(\dot \phi'^{2}-\ddot \phi \phi''\)\nn\,,
\ea
where $Z_n^{\mu\nu}$ is the generalization of the effective metric \eqref{eq:ZuvPX} for the Galileon of order $n$, given  by
\ba
Z_n^{\mu\nu}=(n-1)G_{n,X}(X)\eta^{\mu\nu}+2G_{n,XX}(X)\p^\mu \phi \p^\nu \phi\,.
\ea
The eigenvalues for each  $Z_n{}\mupn$ are now $(n-1)G_{n,X}$ and $(n-1)G_{n,X}+2X G_{n,XX}$ leading to the following expression for the speed of sound for each separate rank of Galileon
\ba
\label{eq:csG}
c^2_n (X) =  \f{(n - 1) G_{n,X} }{(n - 1) G_{n,X} +2X  G_{n,XX} }\,,
\ea
which is a direct generalization of \eqref{eq:csPX} to Galileons. Once again, we can read this as a differential equation for $G_n(X)$ in terms of $c_n(X)$,
\ba
\label{eq:diffG}
\f{\d}{\d X} \log G_{n, X} = \f{n-1}{2X} \mk{\f{1}{c^2_n(X)} -1} \,,
\ea
which is also the direct generalization of \eqref{eq:diffP}.

\subsection{Caustic--free condition}

We are now in measure to check the formation of caustic for simple $SO(p)$--waves in generalized Galileons on flat spacetime. First notice that for every Galileon, the first term in \eqref{eq:eomGal} only contains first order derivatives acting on the field and this first term is thus irrelevant when performing a characteristic analysis. Second, for simple waves the combination $\dot \phi'^{2}-\ddot \phi \phi''$ vanishes identically (see \cite{Babichev:2016hys,Mukohyama:2016ipl}). As a result only the second term in \eqref{eq:eomGal} dictates the formation of caustic in generalized Galileons and all that matters is the light cone dictated by the effective metric $Z_n^{\mu\nu}$, as expected. Interestingly, the effective metric $Z_n^{\mu\nu}$ does not depend\footnote{To be more exact, the dependence in $p$ only enters the conformal factor of the effective metric and hence does not affect the light cone.} on the parameter $p$, and so the theories that avoid caustics are the same independently of that parameter, as far as simple $SO(p)$--waves are concerned.\\

Note that in the equation of motion \eqref{eq:eomGal}, the coefficient of  the term $Z_n^{\mu\nu}$ vanishes for $n\ge p+2$, which means that the characteristic analysis applies only for $2\le n\le p+1$.  This is precisely the reason why  going beyond the planar wave configuration can allow us to study caustics for higher--order Galileon Lagrangians.  As mentioned earlier, so long as one considers planar waves ($p=1$), the caustic--free condition constrains $\L_2$ only (\ie $P(X)$--types of theories).  The $SO(p)$--wave with $p>1$ allows us to constrain $n>2$ Galileon Lagrangians.\\

We can now apply the same analysis as was performed for $P(X)$ theories in section~\ref{sec:charac}.
Focusing again on the high--energy limit and performing the substitution $\p_\mu \to k_\mu$, the relevant solution for the vector $k$ is $k_{\mu}^{(\pm)}=\mk{ \xi_{n, \pm}  ,-1}$ (assuming again  $Z^{00}_n\ne0$) with
\ba
\label{xipmn}
\xi_{n,\pm}=\f{Z_n^{01}\pm \sqrt{-\det Z_n} }{Z_n^{00}}
=\f{v\pm c_n}{1\pm v c_n}\,,
\ea
and once again with $v=-\frac{k_r}{k_t}$
and with the speed of sound now given in Eq.~\eqref{eq:csG}.
The characteristics follows the equations of motion  \eqref{eq:eomGal} with $\xi_{n\pm}$, namely,
\ba
\d k_t + \xi_{n,\pm} \d k_r = 0\,.
\ea\vspace{5pt}

The rest of the analysis is a natural generalization of \cite{Mukohyama:2016ipl}.
Imposing as before the linear dependency, which is the only general condition one can impose that would manifestly prevent the formation of caustics, $k_t = ak_r+b$,
and using the relation  $X=-k_t^2+k_r^2=[(1-a^2)k_r^2-2abk_r-b^2 ]/2$, the speed of sound is given once again in terms of $a$ and $b$ as in \eqref{eq:csab}, namely
\ba
\label{eq:csab2}
c_n^2 (X) = 1+\alpha X\,,
\ea
with the same definition \eqref{eq:defal} of $\alpha$, which can be used in \eqref{eq:diffG} to give an expression for the functions $G_n(X)$ that do not generate caustics in simple $SO(p)$--waves. \\

Once again there are only three relevant cases corresponding to \eqref{eq:P(X) no caustic} that we shall consider in turn:
\begin{itemize}
\item $\alpha=0$ or $(f,g)=(\pm 1,0)$: Pure Galileon.

The first case corresponds to $\alpha=0$ or $g=1-a^2=0$ for any $f=b^2\ne 0$, then we recover a trivial sound speed,  $c_n^2=1$, which corresponds to the standard Galileon that satisfies the shift and Galilean symmetry \eqref{eq:GalileanSym},
\ba \label{noncauG1}
G_n(X)\sim X\,.
\ea
One could of course add a constant contribution to $G_n(X)$ but that contribution would be irrelevant as it would be a total derivative in the action.  Without loss of generality we can consider only $(f,g)=(1,0)$.

\item $\alpha=1$ or $(f,g)=(1,\pm 1)$: DBI--Galileon.

For the second case, assuming $\alpha \ne 0$, \ie $f=b^2\ne 0$ and $g=1-a^2\ne 0$, we can take $c_n^2=1\pm X$ (after appropriate rescaling $\alpha$ into $\phi$) and the differential equation for $G_n$ is simply $\p_X \log G_{n,X}=-(n-1)/2(1\pm X)$, leading to
\be
\label{noncauG2}
G_n(X) =
\begin{cases}
a_n \, (1\pm X)^{\f{3-n}{2}}  , \quad {\rm for}\quad  2\le n \le D+1\ \text{and}\ n\neq 3 , \\
a_n \, \log (1\pm X) , \quad {\rm for}\quad  n=3 \, ,
\end{cases}
\ee
where the overall coefficient $a_n$ is independent on the field and its derivatives.  We shall see that this class is the unique theory that enjoys the relativistic Galilean symmetry \eqref{eq:DBIsym} in section~\ref{sec:PoincareInvariance}, and equivalent to DBI--Galileon introduced in \cite{deRham:2010eu} in Appendix~\ref{app:probe-brane}.

\item $\alpha\to \pm\infty$ or $(f,g)=(0,\pm 1)$: Cuscuta--Galileon.

The third case is $\alpha \to \pm\infty$, \ie $f=b^2=0$ for any $g=1-a^2\ne 0$.  In this case we have infinite speed of sound so long as $X\ne 0$, and obtain
\be
\label{noncauG3}
G_n(X) =
\begin{cases}
a_n \, (\pm X)^{\f{3-n}{2}}  , \quad {\rm for}\quad  2\le n \le D+1\ \text{and}\ n\neq 3 , \\
a_n \, \log (\pm X) , \quad {\rm for}\quad  n=3 \, .
\end{cases}
\ee
This corresponds to the extreme--relativistic Galileon introduced in~\cite{Chagoya:2016inc}, or
{\it cuscuta--Galileon} as it is a natural Galilean generalization of the cuscuton~\cite{Afshordi:2006ad,Afshordi:2007yx,Afshordi:2009tt}.  Indeed, the cuscuta--Galileon has the same property of the cuscuton, \ie the infinite speed of sound and the kinetic term reducible to a total derivative for homogeneous configuration in flat spacetime.  Specifically, for $n=3$,
\be \log X\ \Box\phi \sim \ddot\phi \log \dot\phi \sim \f{d}{dt}(\dot\phi \log\dot\phi) , \ee
and for $n\geq 4$, generalized Galileon Lagrangians identically vanish for homogeneous configuration due to their anti--symmetry.  Therefore, the equation of motion for $\phi$ does not involve $\ddot\phi$ and hence the scalar field is a nondynamical auxiliary field `parasitizing' the dynamics of fields that it couples to.
We shall see that the cuscuta--Galileon is the unique theory that enjoys the extreme--relativistic Galilean symmetry \eqref{eq:exGalsym} in section~\ref{sec:PoincareInvariance}, and equivalent to the model considered in~\cite{Chagoya:2016inc} in Appendix~\ref{app:probe-brane}.\\
\end{itemize}

Even though we have only focused on a specific type of simple waves, the absence of caustic entirely constrained the covariant form of the action.
For the second case with $n=2$, we recover the DBI action, as pointed out in \cite{Mukohyama:2016ipl}.
Actually, there are wider class of theories that satisfy the caustic-free condition.
For the case $\alpha \ne 0$ with $n>2$, we see that the form of the generalized Galileon is very constrained just as was the case of $P(X)$ and, as we shall see below, the corresponding theories are none other the DBI--Galileon that can be obtained from a the Lovelock invariants in higher--dimensional probe brane model.
In addition, for the case $\alpha \to \pm \infty$, similar form of Lagrangians is obtained, which generalizes the cuscuton to Galileon type interactions.  We shall see this cuscuta--Galileon corresponds to an extreme--relativistic limit of the DBI--Galileon~\cite{Chagoya:2016inc}.
Again, it is intriguing to note that all three models enjoy the shift symmetry and an additional global symmetries, which are inherited from a higher--dimensional description.  \\

\subsection{Linear Combination of Generalized Galileons}

When combining multiple generalized Galileons of different rank together
\ba
\L=\sum_{n=2}^{D+1} G_n(X) \mathcal{G}_{n-2}[\Phi]\,,
\ea
the effective metric depends explicitly on $p$ and $\phi'$,
\ba
Z^{\mu\nu}=\(\sum_{n=2}^{D+1}\frac{(n-1)G_{n,X}(X)\(\frac{\phi'}{r}\)^{n-2}}{(p+1-n)!}\)\eta^{\mu\nu}
+2\(\sum_{n=2}^{D+1}\frac{G_{n,XX}(X)\(\frac{\phi'}{r}\)^{n-2}}{(p+1-n)!}\)\p^\mu \phi \p^\nu \phi\hspace{30pt}
\ea
and so does the speed of sound,
\ba
c_s^2=\frac{\sum_n  \frac{(n-1)}{(p+1-n)!}\(\frac{\phi'}{r}\)^n G_{n,X}}{\sum_n  \frac{1}{(p+1-n)!} \(\frac{\phi'}{r}\)^n \((n-1) G_{n,X}+2X G_{n,XX}\)}\,.
\ea
Since generalized Galileons of different rank $n$ lead to contributions with different powers of $\frac{\phi'}{r}$ and different dependencies on $p$, the only way for the speed of sound square to be linear in $X$ is if the  same differential equation
\ba
\frac{(n-1)G_{n,X}}{(n-1) G_{n,X}+2X G_{n,XX}}=1+\alpha X\,,
\ea
be satisfied with exactly the same constant $\alpha$ for every single $G_n(X)$.\\

This implies that even when combining different generalized Galileons together, the absence of caustic for simple $SO(p)$ waves is only guaranteed for the following three theories:
\be
\L_{\rm Gal}= \sum_{n=2}^{D+1} a_n\  X\, \mathcal{G}_{n-2}[\Phi]\,,
\ee
for $\alpha=0$ (where the coefficients $a_n$ are arbitrary constant) which corresponds to the standard `non--relativistic' Galileon introduced in \cite{Nicolis:2008in},
\be
\L_{{\rm DBI-Gal}}= \sum_{n=2}^{D+1} a_n \(1\pm X\)^{(3-n)/2} \mathcal{G}_{n-2}[\Phi]\,,
\ee
for finite $\alpha$ (after rescaling of the field) which corresponds to the DBI--Galileon introduced in \cite{deRham:2010eu}, and
\be
\L_{{\rm cuscuta-Gal}}= \sum_{n=2}^{D+1} a_n\ (\pm X)^{(3-n)/2} \mathcal{G}_{n-2}[\Phi]\,,
\ee
for $\alpha = \pm \infty$ which is the Galilean generalization of the cuscuton, and corresponds to the `extreme--relativistic' Galileon introduced in \cite{Chagoya:2016inc}.
For $n=3$, the term $(1\pm X)^{(3-n)/2}$ and $(\pm X)^{(3-n)/2}$ should be understood as $\log(1\pm X)$ and $\log (\pm X)$.  \\

To summarize, we started here with generalized Galileon Lagrangians in $D=d+1$ flat spacetime dimensions with arbitrary functions $G_n(X)$ that enjoy the shift symmetry~\eqref{eq:ShiftSym}, and imposed the absence of the caustics for simple $SO(p)$--waves.
Interestingly, the caustic--free condition seems to single out the models that are endowed with an additional global symmetry.  Indeed, all the Galileon Lagrangians with $G_n\sim X$ enjoy the global Galilean symmetry given~\eqref{eq:GalileanSym}.  The second class of models with $G_n\sim(1\pm X)^{(3 - n)/2}$ as in \eqref{noncauG2} enjoy the global symmetry provided \eqref{eq:DBIsym} as shown in section~\ref{sec:PoincareInvariance}.  The second class of models with $G_n\sim (\pm X)^{(3 - n)/2}$ as in \eqref{noncauG3} enjoy the global symmetry \eqref{eq:exGalsym} which we shall show in section~\ref{sec:PoincareInvariance}.  The consistency of these types of theories is therefore likely tight with the existence of the additional symmetry.

\section{Higher--dimensional Poincar\'e invariance}
\label{sec:PoincareInvariance}

The previous arguments have singled out a special class of Galileon interactions simply based on the requirement that simple $SO(p)$--waves are free of caustic in flat spacetime\footnote{This is not to say that such models would never develop caustics for other types of waves.}.
In what follows we shall prove that these interactions are actually the unique ones that are invariant under the extreme--/relativistic Galilean transformation reminiscent from higher dimensional Poincar\'e invariance and are therefore equivalent to the cuscuta--/DBI--Galileon interactions.  The equivalence to the original form derived in \cite{Chagoya:2016inc} and \cite{deRham:2010eu}, respectively, shall be explicitly shown in Appendix~\ref{app:probe-brane}.

\subsection{DBI--Galileon global symmetry}
\label{app:sym}

In this section, we shall prove that the specific generalized Galileon interactions found previously
\ba
\label{eq:Galapp}
\L_{n}=G_n(X)\,  \delta^{\mu_1 \cdots \mu_{n-2}}_{\nu_1 \cdots \nu_{n-2}} \, \Phi^{\nu_1}_{\mu_1}\cdots \Phi^{\nu_{n-2}}_{\mu_{n-2}} \,,
\ea
with
\ba
G_n=a_n \( f + g X\)^{\frac{3-n}{2}}\,,
\ea
are the {\it unique operators that are invariant under the relativistic Galilean transformation}
\ba
\phi &\to& \phi+\delta \phi\\
\label{eq:DBIsymapp}
\delta \phi&=&c + f v_\mu x^\mu + g \phi v^\mu \p_\mu \phi\,.
\ea
Without loss of generality, we can normalize $f,g$ and consider three cases: $(f,g)=(0,0),(\pm 1,0),(1,\pm 1),(0,\pm 1)$.  We shall see that these parameters $f,g$ precisely coincide with those appeared in the previous section.\\

The first case $(f,g)=(0,0)$ is the shift symmetry, and the generalized Galileon with arbitrary $G_n(X)$ is the most general scalar tensor theory which enjoys the shift symmetry \cite{Deffayet:2009mn}.
The second case $(f,g)=(\pm 1,0)$ is non--relativistic transformation, and the pure Galileon $G_n=X$ is the unique theory that enjoys the symmetry.
Below we focus on the latter two cases.
In \cite{deRham:2010eu} it was indeed proven that the invariance under the transformation \eqref{eq:DBIsymapp} with $(f,g)=(1,\pm 1)$ is reminiscent to a higher dimensional Poincar\'e invariance in a higher--dimensional bulk.
From this point of view, the first case with $(f,g)=(\pm 1,0)$, which is nothing but the Galileon symmetry, is considered as a non--relativistic case, and the third case with $(f,g)=(0,\pm 1)$ is extreme--relativistic case introduced in \cite{Chagoya:2016inc} (see Table~\ref{Gals}).\\

\begin{table}[t] 
\begin{tabular}{cc|lll}
$f$ & $g$ & Symmetry & Theory & $G_n(X)$ \\ \hline
$0$ & $0$ & Shift symmetry & Generalized Galileon & Arbitrary \\
$\pm 1$ & $0$ & Shift sym.\ \& 5d Non-relativistic & Pure Galileon & $X$ \\
$1$ & $\pm 1$ & Shift sym.\ \& 5d Relativistic & DBI--Galileon & $(1\pm X)^{(3-n)/2}$ \\
$0$ & $\pm 1$ & Shift sym.\ \& 5d Extreme-relativistic & Cuscuta--Galileon & $(\pm X)^{(3-n)/2}$
\end{tabular}
\caption{Theories that enjoy the global symmetry \eqref{eq:DBIsymapp}.  The last three theories are unique theories that enjoy the shift symmetry and the additional symmetries listed.  They are free from the formation of caustic singularity for simple $SO(p)$ waves. }
\label{Gals}
\end{table} 

Under the transformation \eqref{eq:DBIsymapp}, the field and its derivatives transform as follows:
\ba
\delta \(\p_\mu \phi\)&=& f v_\mu + g v^\alpha \p_\mu \(\phi \p_\alpha \phi \)\\
\label{eq:deltaXapp}
\delta X &=& 2v^\alpha \p_\alpha \phi\, (f+gX) + g \phi v^\alpha \p_\alpha X\\
\label{eq:deltaPhiapp}
\delta \Phi\mn &=& g v^\alpha  \p_\alpha \left[\phi \Phi\mn\right] + g v^\alpha \p_\alpha \left[\p_\mu \phi \p_\nu \phi\right]
\,,
\ea
with $X=(\p \phi)^2$.
For concreteness, we first start with the case $n=2$ and then move to arbitrary $n$ (with $2\le n \le D+1$).

\subsection{DBI and cuscuton}

Now let us consider a theory with an arbitrary function $P(X)$
\ba
\L=P(X)\,.
\ea
Under the transformation \eqref{eq:DBIsymapp}, a $P(X)$ model transforms as
\ba
\delta \L&=&2 P'(X) v^\alpha \p_\alpha\phi (f+gX) + g \phi\, v^\alpha \p_\alpha P(X)\\
&=& v^\alpha \p_\alpha \phi \left[2 P'(X)(f+gX)-gP(X)\right]\,,
\label{eq:deltaPXapp}
\ea
where in the last line we ignored total derivatives.
Requiring $\delta \L$ to vanish leads to
\ba
P(X)=\sqrt{f+gX}
\ea
which is precisely the DBI action for $(f,g)=(1,\pm 1)$, and the cuscuton for $(f,g)=(0,\pm 1)$. Therefore, as already well--known, DBI is the unique $P(X)$ model that enjoys the additional global symmetry \eqref{eq:DBIsymapp} with $(f,g)=(1,\pm 1)$.  In addition, we found that the cuscuton is the unique model that enjoys the additional global symmetry \eqref{eq:DBIsymapp} with $(f,g)=(0,\pm 1)$.
Notice that this result is entirely independent of the particular configuration we choose to take (\ie at no point have we assumed an $SO(p)$--symmetry in here) and independent of the number of dimensions.
In what follows we shall show that the DBI--Galileon and the cuscuta--Galileon are the unique single scalar field theories that have no ghost and also enjoy the global symmetry \eqref{eq:DBIsymapp}. \\

\subsection{DBI--Galileon and cuscuta--Galileon}

We now turn to the generalized Galileon interactions \eqref{eq:Galapp}. As shown in \cite{Deffayet:2009mn}, these are the most general single scalar field theories that have no ghost and enjoy the shift symmetry $\phi \to \phi+c$.

\subsubsection{Special Relations}

To simplify the derivation, it will first be convenient to notice that the tensor $X^{\mu\nu}_n$ defined in \eqref{eq:tensorX} satisfies the following relation:
\ba
\label{eq:Xrel}
2 \Phi^{\alpha}_{\mu}\p_\nu \phi X^{\mu\nu}_n=\p_\mu X X^{\mu\alpha}_{n}\,,
\ea
for any $n\ge 0$, where we should not confused the scalar $X=(\p \phi)^2$ with the tensor $X^{\mu\nu}_n$. Indeed the relation \eqref{eq:Xrel} is trivially satisfied for $n=0$ and for $n=1$. Now assuming that the relation \eqref{eq:Xrel} is satisfied at order $n-1$, one can easily show that it is satisfied at order $n$, indeed, using \eqref{eq:recX},
\ba
2 \Phi^{\alpha}_{\mu}\p_\nu \phi X^{\mu\nu}_{n}&=&2 \Phi^{\alpha}_{\mu}\p_\nu \phi \(-n\, \Phi^{\mu}_{\beta}X^{\beta \nu}_{n-1}+\Phi_{\beta\gamma}X^{\beta\gamma}_{n-1}\eta^{\mu\nu}\) \notag\\
&=& -n \Phi^{\alpha}_{ \mu}\p_\beta X X^{\beta\mu}_{n-1}+\Phi_{\beta\gamma}\p^\alpha X X^{\beta \gamma}_{n-1}
=\p_\mu X\,  X^{\mu\alpha}_n \,.
\ea
The relation \eqref{eq:Xrel} is therefore satisfied for all $n$.

\subsubsection{Transformation}

We now consider a generic generalized Galileon \eqref{eq:Lnapp} and apply the transformation \eqref{eq:DBIsymapp},
\ba
\delta \L_n=G_n'(X)\mathcal{G}_{n-2}\delta X + (n-2)G_n(X)X^{\mu\nu}_{n-3} \delta \Phi\mn \,.
\ea
Using the relations provided in \eqref{eq:deltaXapp} and \eqref{eq:deltaPhiapp}, we have
\ba
\label{eq:trans1app}
\delta \L_n&=&\mathcal{G}_{n-2} \left[2 G'_n(X) v^\alpha \p_\alpha \phi\, (f+gX)+g\phi v^\alpha \p_\alpha G_n(X)\right]\\
&+& g G_n(X)\left[(n-2) \mathcal{G}_{n-2} v^\alpha \p_\alpha \phi + \phi v^\alpha  \p_\alpha \mathcal{G}_{n-2}\right]
+ g (n-2)G_n(X) X^{\mu\nu}_{n-3}\, v^\alpha \p_\alpha \left[\p_\mu \phi \p_\nu \phi\right]\nn\,.
\ea
We first notice that the last term is a total derivative, indeed using the relation \eqref{eq:Xrel} we have for any integers $n$ and $m$,
\ba
G_n(X) X^{\mu\nu}_{m}\, v^\alpha \p_\alpha \left[\p_\mu \phi \p_\nu \phi\right]
&=& 2 G_n(X)v_\alpha \Phi^{\alpha}_{ \mu}\p_\nu \phi X^{\mu\nu}_m \nn\\
&=& v_\alpha G_n(X)\p_\mu X X^{\mu \alpha}_m \nn\\
&=& v_\alpha \p_\mu \(\mathbb{G}_n(X)\) X^{\mu \alpha}_m \nn\\
&=& - v_\alpha \mathbb{G}_n(X) \p_\mu X^{\mu \alpha}_m \equiv 0\,,
\ea
where we used the notation $\mathbb{G}'_n(X)=G_n(X)$.\\

Going back to the transformation \eqref{eq:trans1app} and performing an integrations by parts we get
\ba
\delta \L_n=v^\alpha \p_\alpha \phi \, \mathcal{G}_{n-2}\left[2G_n'(X)(f+gX)+g(n-3)G_n(X)\right]\,,
\ea
which precisely matches the result for $n=2$ found in \eqref{eq:deltaPXapp} but is now valid for arbitrary $n$.\\

Requiring the action to be invariant under the transformation \eqref{eq:DBIsymapp} then imposes a very specific form for the functions $G_n(X)$. Indeed since by itself $v^\alpha \p_\alpha \phi \, \mathcal{G}_{n-2}$ is a total derivative, for any constant vector $v^\alpha$, it follows that the combination $\left[2G_n'(X)(f+gX)+g(n-3)G_n(X)\right]$ should be a constant for the action $\int \L_n$ to be invariant. This imposes
\ba
G_n(X)\sim (f+gX)^{\f{3-n}2}\,,
\ea
up to an irrelevant constant (and for $n=3$, it is understood that the solution is logarithmic). This is precisely the form of the function obtained in \eqref{noncauG2} and \eqref{noncauG3} by requiring the absence of caustic formation. \\

This establishes a link between the absence of caustic for simple $SO(p)$--waves and the existence of an additional global symmetry and hence the DBI--Galileon and the cuscuta--Galileon arising from a higher dimensional probe--brane setup.
We can therefore conclude that
the pure Galileon, the DBI--Galileon and the cuscuta--Galileon  are the {\it unique} set of single field interactions which enjoy a shift symmetry $\phi \to \phi+c$ and the additional symmetry inherited from higher dimensional description, have no \Ost ghost {\it and} are manifestly free of caustics as far as simple $SO(p)$--plane waves are concerned.  As mentioned earlier, this is not to say that these models are caustic--free for any type of configurations but it does diagnose a link between the absence of caustic in some configurations and the existence of a global symmetry, or the link with a higher--dimensional Poincar\'e invariance, at least in the case where the shift symmetry is preserved.

\section{Outlook}
\label{sec:Outlook}

In most models with non--standard kinetic term, the formation of caustic in certain configurations is not surprising as the effective metric evolves as a function of the field and its own derivatives. While the onset of caustic can themselves not be trusted, the existence is certainly physically unappealing. For instance models where caustics {\it always inexorably} appear in some physically motivated situations, (for instance during gravitational collapse) leave little to say about themselves. Determining precisely how generic caustics are to form in a given model and whether one can bypass them without fine--tuning of the initial conditions would be an ultimate goal, but in the manuscript we took upon the lesser goal of determining when caustics are guaranteed not to form in specific configurations, namely when dealing with spherical simple waves of arbitrary dimensions (or $SO(p)$ waves, where $p\leq d$ is an arbitrary integer and $d$ is the number of space dimensions). \\

For standard plane waves, it was shown recently in \cite{Babichev:2016hys} that the pure standard kinetic term with no other modifications to the kinetic term was the unique shift--symmetric scalar field model manifestly caustic--free. However when dealing with simple plane waves it was recently shown in \cite{Mukohyama:2016ipl} that the exceptional condition pointed out in \cite{Deser:1998wv} could allow for the DBI scalar field model to avoid caustics. Interestingly the DBI model can be seen as arising from extra dimensions using a probe--brane approximation and enjoys an additional relativistic global Galileon symmetry \cite{deRham:2010eu}. In this manuscript we have solidified this link between the existence of a (non--/extreme--)relativistic global Galileon symmetry and the total absence of caustics for simple spherical (or $SO(p)$) waves of arbitrary dimensions when the shift symmetry is preserved and have re-derived the full `pure' Galileon, the full DBI--Galileon type of interactions, as well as the cuscuta--Galileon as Galilean generalization of the cuscuton, that arise from considering a probe--brane in a Minkowski higher dimensional bulk. These results highlight the link between the higher--dimensional picture and the absence of caustics and could explain why the (DBI--/cuscuta--)Galileon models are special. \\

We emphasize that the scalar field models presented here will still generate caustics in some situations (for instance when relaxing the simple wave configuration, or when considering more generic setups and a curved background). However the absence of caustic for simple spherical waves make them appealing and may hint on some underlying structure that these models preserve. The existence of caustics per se does not invalidate a theory as a whole as it simply indicates that new physics ought to be included to describe the evolution of the system. It would  however be interesting to understand whether these types of theories always leads to caustics unless very specially--tuned initial conditions are considered.

\vspace{0.5cm}
\noindent{\bf Acknowledgments:}
We would like to thank Shinji Mukohyama and Andrew J. Tolley for useful discussions.
While at CWRU, CdR was supported by a Department of Energy grant DE-SC0009946.
At ICL, CdR is supported by a Royal Society Wolfson Merit Award.
HM was supported by the Kavli Institute for Cosmological Physics at the University of Chicago through grants NSF PHY-0114422 and NSF PHY-0551142 and an endowment from the Kavli Foundation and its founder Fred Kavli.
At IFIC, HM is supported by MINECO Grant SEV-2014-0398,
PROMETEO II/2014/050,
Spanish Grant FPA2014-57816-P of the MINECO, and
European Union’s Horizon 2020 research and innovation programme under the Marie Sk\l{}odowska-Curie grant agreements No.~690575 and 674896.

\appendix

\renewcommand{\thesubsection}{\Alph{section}.\arabic{subsection}}

\section{Probe--brane in higher dimensional Minkowski}
\label{app:probe-brane}

When introduced in \cite{deRham:2010eu} from the probe--brane approximation in higher--dimensional Minkowski, the DBI--Galileon operators were presented in a slightly different form. Also, the cuscuta--Galileon had similar form when derived from probe brane in \cite{Chagoya:2016inc}.
To confirm that we are indeed dealing with exactly the same objects, we show here the relation between their respective expressions.

In the picture of \cite{deRham:2010eu}, we consider a brane localized at $y\sim\phi(x^\mu)$ embedded within a Minkowski five dimensional bulk. The probe--brane approximation assumes that the backreaction of the brane on the five--dimensional geometry is negligible and we can thus keep treating the five--dimensional bulk as Minkowski even though the brane may carry a tension $\lambda$ and an Einstein--Hilbert term $R$. In addition to these contributions localized on the brane, the five dimensional bulk carries a five--dimensional curvature term and even possible a five-dimensional Gauss--Bonnet term. These five--dimensional contributions lead to  Gibbon--Hawking boundary terms on the brane, which are noting other than the trace of the extrinsic curvature $K$ for the five-dimensional Einstein Hilbert term and a more complicated version for the Gauss--Bonnet term which involves cubic order in the extrinsic curvature. The induced metric  on the brane can be written as
\ba
g\mn=f^{1/4}\(\eta\mn+\frac{g}{f}\p_\mu \phi \p_\nu \phi\)\,,
\ea
where the dimensionless coefficients $f$ and $g$ have been introduced to better compare with the theories we have derived in this paper. In terms of this induced metric, we can infer the contributions to the brane action from the tension $\lambda$ on the brane, which we denote as $S_\lambda$; the Gibbon--Hawking boundary term on the brane associated with the five--dimensional Einstein--Hilbert term, which we denote as $S_K$; the induced Einstein--Hilbert term on the brane which we denote as $S_R$ and finally the Gibbon--Hawking boundary term on the brane associated with the five--dimensional Gauss--Bonnet term, which we denote as $S_K$. Those take the following expressions in terms of $\phi$ \cite{deRham:2010eu},
\begin{align}
S_\lambda &\sim \int \d^4 x \sqrt{f+gX} ,\notag\\
S_K &\sim \int \d^4 x \mk{ [\Phi] - \f{[\Sigma]}{f+gX} } ,\notag\\
S_R &\sim \int \d^4 x \mk{ \f{[\Phi]^2-[\Phi^2]}{(f+gX)^{1/2}} + \f{2([\Sigma^2]-[\Phi][\Sigma])}{(f+gX)^{3/2}} } ,\notag\\
S_{\rm GB} &\sim \int \d^4 x \left( \f{1}{f+gX}([\Phi]^3+2[\Phi^3]-3[\Phi^2][\Phi]) \right. \notag\\
&\hspace{2cm} \left.+ \f{3}{(f+gX)^2} \mk{2 [\Phi][\Sigma^2] - 2 [\Phi^3]- [\Phi][\Sigma]^2 + [\Phi][\Sigma^2] } \right) \,,
\end{align}
where we have use the notation $X=(\p \phi)^2$, $\Phi\mn=\p_\mu \phi \p_\nu \phi$, $\Sigma^n\mn= \p_\mu \phi \p_\alpha \phi \Phi^n{}^\alpha_\nu$ and where square brackets represent traces of tensors. All the raising and lowering of indices is taken with respect to the four--dimensional Minkowski metric $\eta\mn$. 

Below we show these invariants are indeed equivalent to generalized Galileon with \eqref{noncauG2} and \eqref{noncauG3}, which as well as the pure Galileon are obtained as the unique set that is free from caustics as far as simple $SO(p)$ waves are concerned which were given in section~\ref{sec:GeneralizedGalileons} by
\ba
\L_n= G_n(X) \mathcal{G}_{n-2}[\Phi]\,, \quad{\rm for }\quad n\ge 2\,,
\ea
with 
\ba
\mathcal{G}_k[\Phi]= \delta^{\mu_1 \cdots \mu_k}_{\nu_1 \cdots \nu_k} \Phi^{\nu_1}_{\mu_1}\cdots \Phi^{\nu_k}_{\mu_k}\,,
\ea
and to satisfy the caustic--free conditions for simple $SO(p)$ wave configurations, it was shown in section \ref{sec:GeneralizedGalileons} that the functions $G_n(X)$ had to take the following form  (see Eqns.~ \eqref{noncauG1}, \eqref{noncauG2} and \eqref{noncauG3}),
\ba
G_n(X)\sim  (f+g X)^{(3-n)/2}\,, \quad {\rm for}~~ g \ne 0\,,
\ea
where the $\log$ is understood for $n=3$ and $G_n(X)=X$ for $g=0$. 

Focusing first on the case where $g\ne0$, for $S_\lambda$, we directly see that $S_\lambda$ is equivalent to the caustic--free Lagrangian derived in \eqref{noncauG1}, \eqref{noncauG2} and \eqref{noncauG3} depending on the respective values of $f$ and $g$ for $n=2$.

For $S_K$, we see the following correspondence (ignoring the boundary terms in four dimensions),
\ba
S_K&\sim&\int \d^4x \(\Box \phi - \frac{\p_\mu\p_\nu \phi \p^\mu \phi \p^\nu \phi}{f+gX}\)
\sim \int \d^4x \frac{\p_\mu X \p^\mu \phi}{f+gX} \nn \\
&\sim& \int \d^4x \p_\mu \log (f+gX) \p^\mu \phi
\sim \int \d^4x \log (f+gX) \mathcal{G}_1[\Phi]\,.
\ea
For $S_R$, it is first convenient to notice that
\ba
\int \d^4 x \frac{[\Phi]^2-[\Phi^2]}{(f+gX)^{1/2}}
=\int \d^4 x\frac{[\Phi][\Sigma]-[\Sigma^2]}{(f+gX)^{3/2}}\,,
\ea
From these relations we can easily show that $S_R$ corresponds to the same operator found from the caustic--free condition,
\ba
S_R&\sim&\int \d^4 x\(\frac{[\Phi]^2-[\Phi^2]}{(f+gX)^{1/2}} +2 \frac{[\Sigma^2]-[\Phi][\Sigma] }{(f+gX)^{3/2}}\)\nn\\
&\sim&\int \d^4 x \, \frac{[\Phi]^2-[\Phi^2]}{(f+gX)^{1/2}} \sim \int \d^4 x \(f+gX\)^{-1/2} \mathcal{G}_2[\Phi]\,.
\ea
The same goes through for the term $S_{\rm GB}$ after noticing that
\ba
\int \d ^4 x \frac{\mathcal{G}_3[\Phi]}{f+gX} \sim \int \d ^4 x \frac{1}{(f+gX)^2}\(2[\Phi][\Sigma^2]-2[\Sigma^3]-[\Phi]^2[\Sigma]+[\Phi^2][\Sigma]\)\,,
\ea
so that the term $S_{\rm GB}$ can be written as
\ba
S_{\rm GB}\sim \int \d^4 x\ \frac{\mathcal{G}_3[\Phi]}{f+gX}\,,
\ea
hence matching the relation \eqref{noncauG2} and \eqref{noncauG3} for $n=5$.  We could of course reproduce the procedure in arbitrary dimensions and recover all the operators found in \eqref{noncauG2} and \eqref{noncauG3} by allowing the higher order Lovelock invariant both on the brane and in the bulk, with their respective boundary terms on the brane.  It is clear that higher Lovelock invariants come along with more powers of curvatures, \ie higher powers of $\Phi\mn$ as well as more powers of inverse metric, \ie higher powers  $(f+gX)^{-1/2}$ leading to precisely the correct scaling in terms of $(f+gX)^{(3-n)/2}$ as found in \eqref{noncauG2} and \eqref{noncauG3}.

The case where $g\to 0$ can be understood as the non--relativistic limit of the previous DBI--Galileon Lagrangians and were shown in \cite{deRham:2010eu} to lead to the standard pure Galileons, which is precisely what was found in  \eqref{noncauG1} when $g=0$.

\bibliographystyle{JHEPmodplain}
\bibliography{refs}

\end{document}